\documentclass[twocolumn,showpacs,amsmath,amssymb,nature]{revtex4-1}
\usepackage{natbib}
\usepackage{graphicx}
\usepackage{dcolumn}
\usepackage[english]{babel}
\usepackage{amsmath}
\usepackage{amssymb}
\usepackage{tabularx}
\usepackage{multirow}
\usepackage[all]{xy}
\usepackage{appendix}
\usepackage{xspace}
\usepackage{color}
\setcitestyle{super}

\newcommand{\ket}[1]{\ensuremath{|#1\rangle}\xspace}

\begin{document}
\title{Subcycle engineering of laser filamentation in gas by harmonic seeding}

\author{P. B\'ejot} \email{pierre.bejot@u-bourgogne.fr}
\author{G. Karras}
\author{F. Billard}
\author{J. Doussot}
\author{E. Hertz}
\author{B. Lavorel}
\author{O. Faucher}

\affiliation{Laboratoire Interdisciplinaire CARNOT de Bourgogne, UMR 6303 CNRS-Universit\'e de Bourgogne, BP 47870, 21078 Dijon, France.}

\maketitle

\begin{figure}[htbp!]
  \begin{center}
    \includegraphics[keepaspectratio, width=8.6cm]{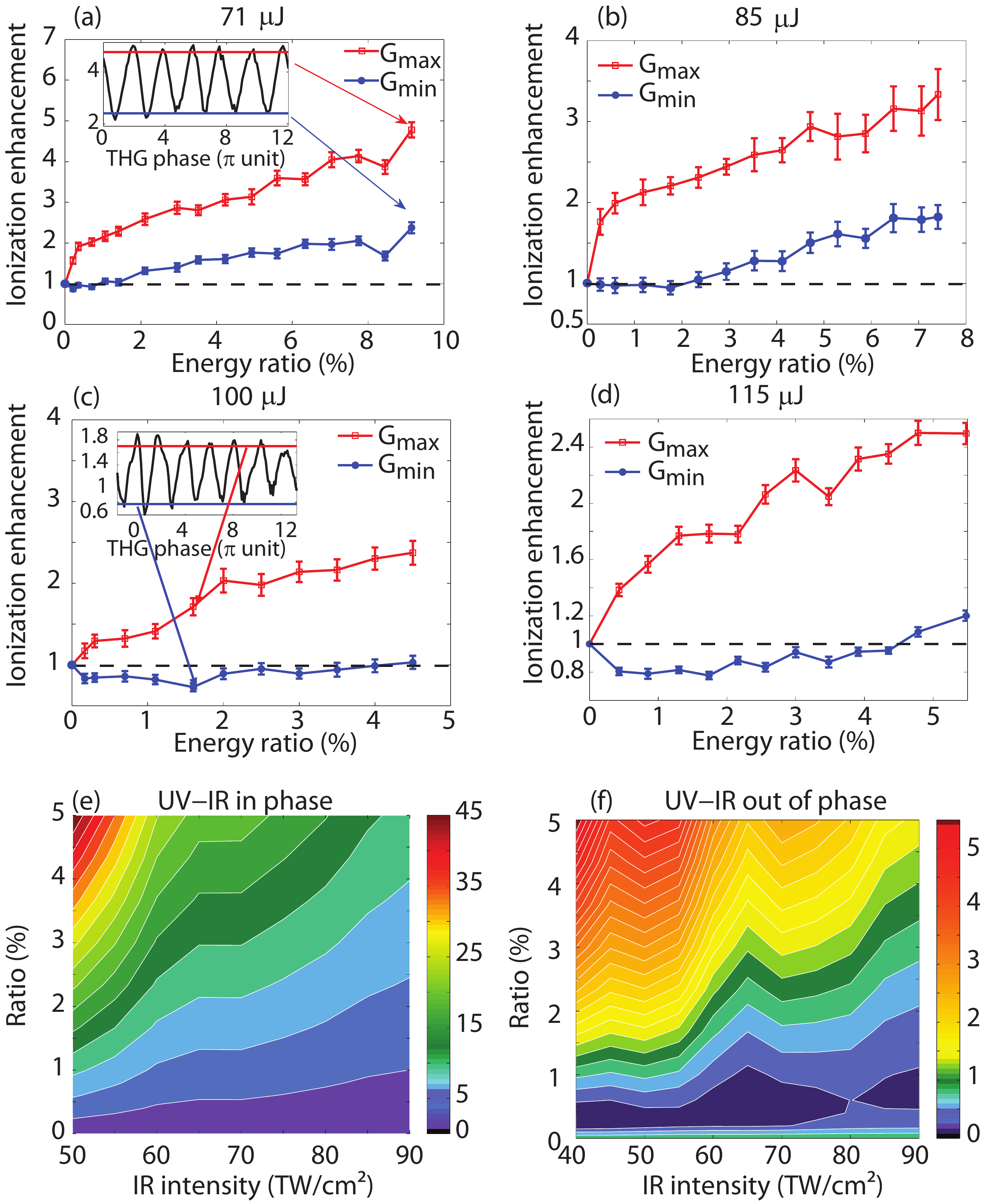}
  \end{center}
  \caption{\textbf{Modification of the nonlinear optical properties of argon by harmonic seeding.} \textbf{a-d}, Ionization yields as a function of the UV energy (expressed in \% of the energy of the IR pulse) for different IR energies $E_{\textrm{IR}}$ (\textbf{a}) 71\,$\mu$J, (\textbf{b}) 85\,$\mu$J, (\textbf{c}) 100\,$\mu$J, and (\textbf{d}) 115\,$\mu$J. The peak intensities $I_{\textrm{IR}}$ (expressed in TW/cm$^2$) can be approximated by $I_{\textrm{IR}}\simeq0.5E_{\textrm{IR}}$, where $E_{\textrm{IR}}$ is expressed in $\mu$J. The red (blue) curves correspond to the relative phase maximizing (minimizing) the ionization yield. The insets show the experimental ionization yield modification as a function of the relative phase between the two electric fields. For each point, the relative phase between pump and UV beams has been varied over 12 $\pi$. The error bars correspond to one standard deviation of the ionization yield modification measured over the full relative phase range. \textbf{e,f}, Calculated ionization yields as a function both IR and UV intensities for constructive (\textbf{e}) and destructive (\textbf{f}) interferences. The calculations have been performed with a 20 cycles laser pulse in order to reduce the computation time.}
  \label{Results1}
\end{figure}

\textbf{Manipulating at will the propagation dynamics of high power laser pulses is a long-standing dream whose  accomplishment would lead to the control of a plethora of fascinating physical phenomena emerging from  laser-matter interaction. The present work represents a significant step towards such an ideal control by manipulating the nonlinear optical properties of the gas medium at the quantum level. This is accomplished by engineering the intense laser pulse experiencing filamentation at the subcycle level with a relatively weak ($\simeq$1$\%$) third-harmonic radiation. The control results from quantum interferences between a single and a two-color (mixing the fundamental frequency with its 3rd harmonic) ionization channel. This mechanism, which depends on the relative phase between the two electric fields, is responsible for wide refractive index modifications in relation with significant enhancement or suppression of the ionization rate. As a first application, we demonstrate the production and control of an axially modulated plasma channel that could be used for quasi-phase-matched laser wakefield acceleration.}\linebreak
\indent Since its first observation\cite{Braun95} in gases by Braun \emph{et al.} in 1995, the nonlinear propagation of ultrashort ultra-intense laser pulses, \emph{i.e.} filamentation, has attracted a lot of interest in recent years because of its physical interest\cite{ChinReport,MysyReport,BergeReport,KasparianW08}, as well as its important applications including terahertz\cite{Damico} and supercontinuum\cite{AkozbekWL} generation, remote sensing\cite{KasparianScience}, attosecond\cite{CouaironAtto} and high-harmonics\cite{SteingrubeHHG} physics, spectroscopy\cite{StelmaszczykSpectro}, machining\cite{Kiselev} and lightning protection\cite{KasparianLightning}. The main feature of a filament is its ability to sustain very high intensities (around 50\,TW/cm$^2$) over very long distances in contrast with the predictions of linear propagation theory. Controlling at will the natural characteristics of a filament and its by-products (such as, for instance, the plasma channel left in its wake)  by means of a single control parameter would make the filamentation process an even more versatile tool  for applications. First attempts devoted to control the filament characteristics were realized by using a temporal \cite{Ackermann} or a spatial \cite{SpatialControlFiLament1,SpatialControlFiLament2,SpatialControlFiLament3} pulse shaper eventually coupled with a closed-loop algorithm. Such methods, based on the engineering of the pulse envelope, successfully controlled either the spectral broadening or the plasma channel position. Another way to control the filamentation process based on molecular alignment was also reported\cite{MilchbergAlign}. By manipulating the rotational degree of freedom of molecules with a strong laser pulse, it was shown that the filament length, continuity, and electron density can be manipulated. More recently, it was shown that an energetic Bessel beam co-propagating with a filament can extend by an order of magnitude the length of the latter \cite{MoloRefuelled} by continuously refueling it all along its propagation. Based on a previous proposal\cite{Bejot14}, the current work demonstrates that the properties of a filament generated in a gas can be manipulated by controlling the nonlinear optical properties of the medium at the quantum level. The underlying idea relies on a subcycle engineering technique originally used for sub-femtosecond spectroscopy\cite{wirth} and for controlling high-harmonic generation\cite{Baltuska}. By seeding the filament with a weak third-harmonic pulse, one can engineer the former at the optical cycle scale. By adjusting the phase of the harmonic field, used as the control parameter, one can alter the ionization yield and the refractive index so as to modulate the attributes of the filament. The technique is first implemented in order to apply a control over the length of the filament and its generated supercontinuum. Then, we demonstrate the production and the control of an axially modulated plasma channel. Such a medium, with tunable periodical optical properties, could be used, inter alia, for quasi-phase-matched laser wakefield acceleration or for engineering the temporal trajectories of Airy light bullets. Finally, beyond the control of the nonlinear propagation from microscopic to macroscopic extent by means of a single control parameter, this work casts doubts on the standard model used in the description of the filamentation process. \linebreak
\indent An atom interacting with a non-resonant intense ultrashort laser pulse can simultaneously absorb a large number of photons, leading to its excitation or eventually to its ionization where a fraction of the bound electronic wavepacket is promoted into the continuum. This electronically excited atom exhibits different optical properties as compared to the initial atom, leading to the modification of the propagation of the laser pulse. In a first experiment, we investigate up to what extend the engineering, at the subcycle scale, of an intense IR-pulse with a weak third harmonic beam can modify the optical properties of a gas.
\begin{figure*}
  \begin{center}
    \includegraphics[keepaspectratio, width=17.2cm]{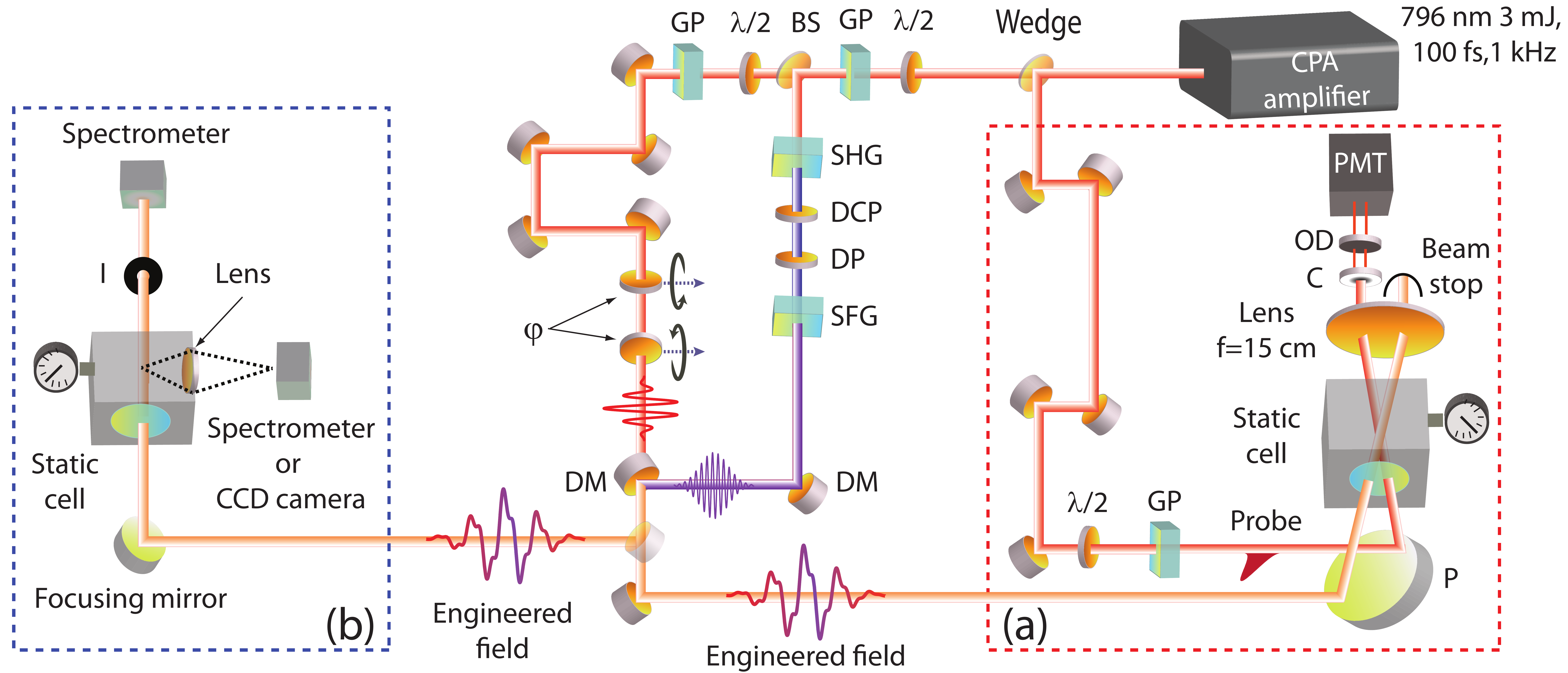}
  \end{center}
  \caption{\textbf{Experimental realization of the eletric field engineering.} Setup for laser-induced cross-defocusing measurements (\textbf{a}) and for the coherent manipulation of a filament produced in argon (\textbf{b}). BS: Beam Splitter, GP: Glan Polarizer, OD: Neutral Optical Density, PMT: Photo Multiplier Tube, DM: Dichroic mirror, C: Coronagraph, DCP: Delay Compensation Plate, DP: Dual Plate ($\lambda$/2 @796\,nm, $\lambda$ @398\,nm), P: Parabolic mirror, I: Iris, SHG: Second-Harmonic Generation, SFG: Sum-Frequency Generation.}
  \label{Setups}
\end{figure*}
The experimental setup used to measure the nonlinear properties of argon is depicted in Fig.\,\ref{Setups}a. The optical source is a 1\,kHz amplified femtosecond laser delivering horizontally polarized, 3\,mJ, 100\,fs pulses at $\lambda_0$=796\,nm. The refractive index change induced by the  IR pump-beam in a static cell filled with argon at 0.5\,bar is measured by the pump-probe cross-defocusing technique\cite{RenardDefocusing,LoriotDefoc}. All the beams are focused with a $f$=15\,cm off-axis parabolic aluminium mirror to avoid both longitudinal and lateral chromatism. A third beam with central wavelength $\lambda_{\textrm{UV}}=\lambda_0/3$ is generated by sum-frequency generation between the fundamental and a second-harmonic pulses itself generated by second-harmonic generation. It is spatially superimposed with the IR pulse by means of a dichroic mirror. The temporal synchronization of the IR and UV pulses is accomplished with the help of a delay line. The polarization of all beams is set vertical with half-wave plates. The cross-defocusing technique relies on the fact that the pump modifies the propagation of the probe pulse by inducing a spatial change on the local refractive index. After the cell, a coronagraph is inserted in the probe beam path. When the pump beam is switched off, the coronagraph obstructs the probe. On the contrary, if the pump beam induces a local refractive index modification, the probe beam size increases in the far field so that a small amount can propagate around the coronagraph. The remaining part of the probe beam is redirected to a photomultiplier tube. One can show that the defocusing signal is proportional to $\Delta n^2$, \emph{i.e.}, the square of the peak to valley change of refractive index experienced by the probe beam\cite{RenardDefocusing}. When the pump pulse precedes the arrival of the probe, the cross-defocusing signal is proportional to the square of refractive index change resulting from the ionization mechanism and accordingly proportional to the square of the amount of free electrons generated by the pump\cite{LoriotDefoc}. It then allows a direct experimental measurement of the latter. The subcycle engineering of the pump pulse is realized by controlling the relative phase between the pump and the UV pulses. Such a phase control is accomplished by inserting into the pump optical path two fused silica plates that rotate symmetrically with respect to a plane perpendicular to the propagation direction. Rotating the two plates then slightly modulates the optical path length and accordingly the relative phase  $\varphi$ between the two pulses, leaving the beam pointing direction unchanged.\linebreak
\textbf{Results}\linebreak
\indent Figures\,\ref{Results1}a-d show the modification of the ionization yields induced by the presence of the UV. The red (resp. blue) curves depict the ionization as a function of the UV beam energy when the two electric fields are in phase (resp. out-of-phase). The insets in Figs.\,\ref{Results1}a,c display the ionization yields as a function of $\varphi$. While at low energy (Figs.\,\ref{Results1}a,b), the ionization yield increases for any phase, at higher energies (Figs.\,\ref{Results1}c,d), just adding less than  2\,\% of UV leads to a decrease (up to 50\,\%) when the two electric fields are out-of-phase. Moreover, the maximal gain on the ionization yield is higher at lower IR intensity. This behavior is in qualitative agreement with the theoretical predictions shown in Figs.\,\ref{Results1}e,f that depict the ionization yield as a function of both pump and UV intensities for in-phase (Fig.\,\ref{Results1}e) and out-of-phase electric fields (Fig.\,\ref{Results1}f). The theoretical results have been obtained by solving the three dimensional time-dependent Schr\"{o}dinger equation capturing the quantum dynamics of argon in the single active electron approximation. As recently theoretically studied\cite{Bejot14}, the modulation of the ionization yield observed in Fig.\,\ref{Results1} results from quantum interferences between two ionization channels: one involving only IR photons and the other mixing IR and a single-UV photons. The confirmation of this effect opens new possibilities for controlling the propagation dynamics of ultra-intense laser pulses and, consequently, all the underlying applications deriving from it.
\linebreak
\indent In order to explore such a possibility, we performed a second experiment where a filament co-propagates with a weak UV beam. The experimental setup is shown in Fig.\,\ref{Setups}b. An 1\,mJ 100\,fs IR beam is focused together with a 30\,$\mu$J UV pulse in the static cell filled with argon. Then, the spectrum of the filament that has experienced a strong broadening due to nonlinear propagation is recorded. In order to avoid the saturation of the spectrometer, the spectral region lying between 750 and 850\,nm is filtered out with a bandpass filter. At the same time, the fluorescence of the plasma channel is imaged by the side of the cell with the help of a $f$=5\,cm lens placed at a distance 2$f$ from the plasma channel and a charge-coupled device camera. In order to confirm that the signal collected by the camera was only due to fluorescence and not to laser scattering, a spectrometer was first put in place of the camera. The camera and the two spectrometers are synchronously triggered with the stepper motor used to rotate the two fused silica plates, allowing to record the laser spectrum, the plasma channel, and the gas fluorescence spectrum as a function of the relative phase between the two electric fields. One has to emphasize that, because of the optical refractive index dispersion, the UV and IR electric fields travel with different phase velocities. As a consequence, the relative phase between the two pulses does not remain constant all along the propagation.
More particularly, one can define the rephasing length $l_{\textrm{reph}}$ as
\begin{equation}
l_{\textrm{reph}}=\lambda_{\textrm{UV}}\left(\frac{1}{n(\lambda_{\textrm{UV}})-n(\lambda_{0})}\right)
\end{equation}
where n($\lambda_0$) (resp. n($\lambda_{\textrm{UV}}$)) is the refractive index of the gas at the IR (resp. UV) laser central wavelength. After a propagation over a distance $l_{\textrm{reph}}$, $\varphi$ increases by 2$\pi$. For a pressure of 1\,bar, $l_{\textrm{reph}}\simeq1\,$cm. In order to explore the dephasing effect, experiments have been performed with both tightly ($f$=15\,cm) and loosely ($f$=50\,cm) focused beams. In the former (resp. latter) case, the filament is about 2\,mm (resp. 6\,cm) long, and is consequently shorter (resp. longer) than $l_{\textrm{reph}}$.\linebreak
\begin{figure*}
  \begin{center}
    \includegraphics[keepaspectratio, width=\textwidth]{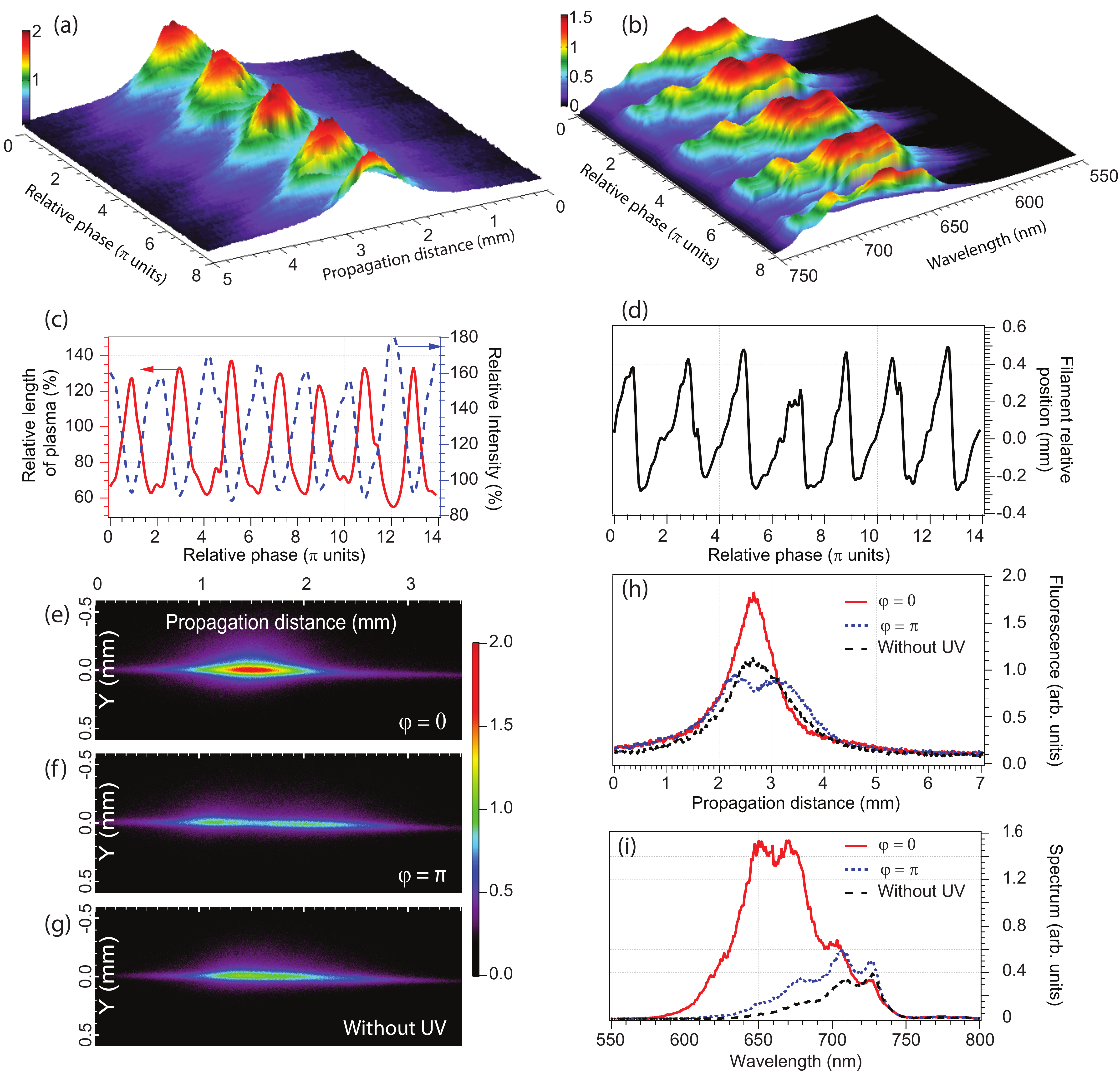}
  \end{center}
  \caption{\textbf{Control of a filament by subcycle engineering in the tightly focused regime.} \textbf{a}-\textbf{d}, Modification of the filament properties as a function $\varphi$. \textbf{a}, Longitudinal profile of the plasma column. \textbf{b}, Supercontinuum generation. \textbf{c}, Plasma column length (solid red) and fluorescence (dashed blue). \textbf{d}, Filament position. \textbf{e-g}, Image of the filament for constructive (\textbf{e}) and destructive (\textbf{f}) quantum interferences between the different ionization channels compared to the case where the filament propagates without the control UV pulse (\textbf{g}). The corresponding longitudinal profiles and spectra are shown in (\textbf{h}) and (\textbf{i}). The pressure is 1.5\,bar.}
  \label{Results2}
\end{figure*}
\indent Figure\,\ref{Results2}a (resp. figure\,\ref{Results2}b) displays the longitudinal profile of the plasma channel (resp. the laser spectrum) as a function of the relative phase between the two electric fields obtained in the tight focusing geometry. In this regime, the relative phase between the two electric fields remains almost constant over the filament length (about 2\,mm).  Depending on the $\varphi$-value, the filament and the plasma left in its wake experience a strong reshaping. In particular, the supercontinuum generated during the filamentation process gets broader when the harmonic field is in phase with the IR field. As shown in Fig.\,\ref{Results2}c, the length and density of the plasma channel strongly depend on $\varphi$. When the two electric fields are in phase (resp. out of phase), the plasma channel is about 40\,\% shorter (resp. longer) as compared to the IR field alone. Modulation of the plasma length is attended by an 80\% increase (or a 10\% decrease) of the maximal plasma density depending on the applied relative phase. Moreover, the filament position depends also on $\varphi$, as depicted in Fig.\,\ref{Results2}d. The two extremal cases that correspond to constructive ($\varphi$=0) and destructive ($\varphi$=$\pi$) quantum interferences are shown in more detail in Figs.\,\ref{Results2}e-i. When $\varphi$=0 (resp. $\varphi$=$\pi$), the filament is more intense and shorter (resp. less intense and longer with a double humps structure) than when the UV is switched off.\linebreak
\begin{figure}
  \begin{center}
    \includegraphics[keepaspectratio, width=8.6cm]{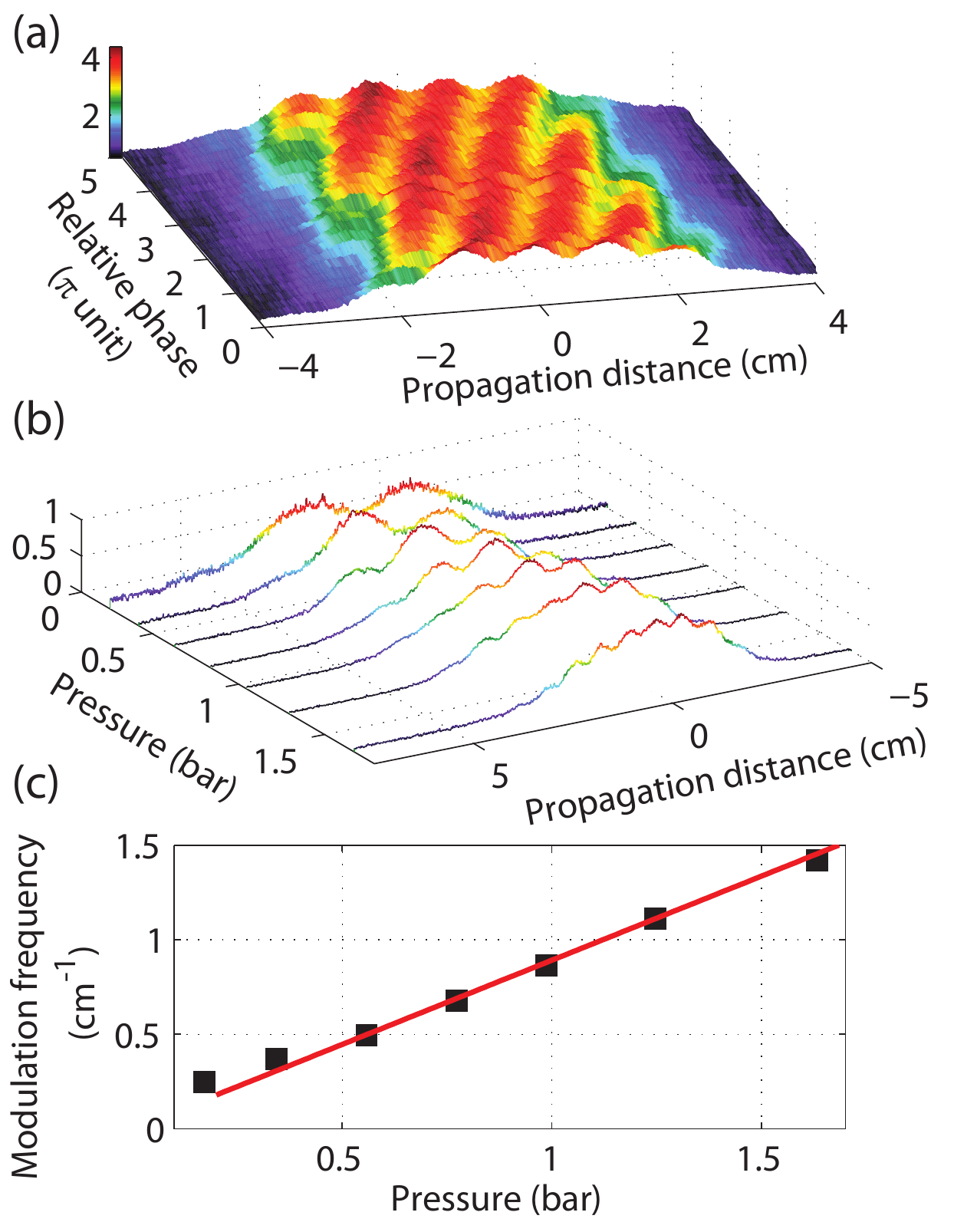}
  \end{center}
  \caption{\textbf{Production and control of an axially modulated plasma channel in the loose focusing regime.} \textbf{a-b}, Longitudinal profile of the plasma channel as a function of $\varphi$ for a pressure of 1\,bar \textbf{(a)} and as a function of the pressure \textbf{(b)}. \textbf{(c)}, Experimental (black squares) and theoretical (red solid line) modulation frequency of the plasma channel as a function of the pressure.}
  \label{Results3}
\end{figure}
\indent Figure\,\ref{Results3}a shows the plasma channel profile along the propagation direction as a function of $\varphi$ obtained in the loose focusing geometry at ambient pressure and temperature conditions. In this case, $\varphi$ does not remain constant over the whole filament length (about 6\,cm). The two electric fields experience several periodic rephasing/dephasing cycles, with the period determined by the difference between their respective phase velocities. As a consequence, the plasma channel is axially modulated with a modulation period of $l_{\textrm{reph}}\simeq$1\,cm. By tuning the initial relative phase between the two electric fields, the plasma channel extrema position continuously shifts along the propagation distance. As depicted in Fig.\,\ref{Results3}b, the period can be controlled by adjusting the pressure, \textit{i.e.}, by changing the phase velocity mismatch between the two electric fields. As shown in Fig.\,\ref{Results3}c, the modulation frequency increases linearly with pressure. The production and the control of an axially modulated plasma channel is promising for applications. For instance, the possibility to create a medium with tunable periodical optical properties could be used, inter alia, for quasi-phase-matched laser wakefield acceleration\cite{Milchberg} without the use of an external waveguide, for engineering the temporal trajectories of Airy light bullets\cite{Zeng}, or for micromachining various periodic optical elements such as waveguides or gratings in the bulk of dielectric media\cite{grating}.\linebreak
\textbf{Discussion}\linebreak
\indent Simulating laser pulse propagation over macroscopic distances in a medium undergoing ionization is complicated by the need to include quantum-mechanical laser-atom dynamics. While ionization yield calculations in atoms are routinely performed on the framework of the time-dependent Schr\"{o}dinger equation, its consideration in the context of two-dimensional laser propagation requires prohibitive numerical resources. For this reason, standard filamentation codes approximate the ionization yield using a simplified analytic formula\cite{PPT} originally developed for purely monochromatic laser fields. Additionally, propagation models accounting for both fundamental and third-harmonic radiations so far neglect the interferences occurring between the different ionization channels \cite{BergeTHG,AkozbekTHG,MysyTHG,CouaironTHG,MoloScattering}. Besides the experimental demonstration showing that the quantum control of the filamentation process can be performed by seeding the filament with a weak harmonic beam, this work then questions the current theoretical framework on which filamentation simulation is based. Specifically, the present results clearly go beyond the validity range of the theoretical models used for describing the filamentation process and highlight that neglecting the interference effect is not valid even when the UV intensity is only a few percent of the fundamental one. Moreover, while the present experimental results have been obtained by externally seeding the filament with a third-harmonic beam, it is known that the latter is generated at the percent level during the propagation of a filament\cite{BergeTHG,AkozbekTHG}. A rational question arising from the present work is up to what extent the self-induced third-harmonic can impact the propagation dynamics of a filament by the exhibited UV-IR interference effect. This question will find an answer only after an accurate model of two-color ionization will be available. We hope this work will stimulate theoretical developments toward this direction.\linebreak
\indent In summary, we have experimentally and theoretically demonstrated that the nonlinear optical properties of a gas experienced by a strong ultrashort laser pulse can be manipulated by a subcycle engineering of the latter. The control has been realized by adding a realistically weak third-harmonic beam that propagates together with the intense laser pulse. Because of quantum interferences occurring during the ionization process, the ionization yield can be either suppressed or enhanced depending on the relative phase between the two electric fields. We have applied this concept to the control of the nonlinear propagation dynamics of a strong laser beam experiencing filamentation. More particularly, we have succeeded in manipulating the supercontinuum generation and the plasma column properties, such as its length and amplitude by adjusting the relative phase between the two electric fields. Moreover, taking advantage of the phase velocity mismatch between the two electric fields, we have created and controlled a sinus-like plasma channel that could be used for quasi-phase-matched laser wakefield acceleration. Demonstrated in argon gas, the control of filamentation by field engineering is based on a non-resonant quantum process. As such, it is a versatile technique that could be applicable to any gas and even to be extended to bulk materials.\linebreak
\textbf{Methods}\linebreak
\footnotesize{\textbf{Phase control by two fused silica plates.}
Considering an incidence angle $\theta$, the optical delay $\tau$ induced by inserting two plates of thickness $e$ and refractive index $n$ in the optical path of a pulse is given by:
\begin{equation}
\tau=\frac{2e}{c}\frac{n-\cos(\theta-\arcsin(\frac{\sin\theta}{n}))}{\sqrt{1-\sin^2\theta/n^2}}.
\end{equation}
A rotation $d\theta$ of  the two plates around  $\theta_0$ induces a relative delay $d\tau$ given by :
\begin{equation}
d\tau\simeq\frac{\partial \tau}{\partial\theta}|_{\theta_{0}}\cdot d\theta+\frac{1}{2}\frac{\partial^2 \tau}{\partial\theta^2}|_{\theta_{0}}\cdot d\theta^2,
\end{equation}

with
\begin{equation}
\frac{\partial \tau}{\partial\theta}= \frac{2e}{c}\sin\theta\left(1-\frac{\cos\theta}{n \sqrt{1-\sin^2\theta/n^2}}\right).
\end{equation}

Around $\theta=0$, the delay is a slowly varying nonlinear function of the rotation angle, whereas at large angle it  is almost linear with $\theta$ but with less resolution. Assuming that the angular resolution given by the minimal step of the  motor is about 4.5 10$^{-5}$ radian, tilting  the angle between  10 and 20$^\circ$  could allow a temporal sampling of 30 points per third-harmonic optical cycle, \emph{i.e.} an 30\,as resolution in the relative delay between the IR and UV pulses. In order to calibrate the phase change induced by rotating the plates, we inserted them into an  HeNe laser-based Michelson interferometer. The interference pattern was then analyzed as a function of the rotation of the plates which provided a direct calibration of the phase control setup.\\

\textbf{Numerical methods.}
The population promoted into the continuum after the interaction was evaluated by solving the time-dependent Schr\"{o}dinger equation\cite{BejotPRL3}. Calculations were performed in argon under the single-active electron approximation. Within the dipole and single active electron approximations, the three dimensional time-dependent Schr\"{o}dinger equation describing the evolution of the electron wavefunction $\ket{\psi}$ in the presence of an electric field $\textbf{E}(t)$ reads:
\begin{equation}
i\frac{d\ket{\psi}}{dt}=(H_0+H_{\textrm{int}})\ket{\psi},
\end{equation}
where $H_0=\mathbf{\nabla}^2/2+V_{\textrm{eff}}$ is the atom Hamiltonian, $H_{\textrm{int}}=\textbf{A}(t)\cdot\boldsymbol{\pi}$, where $\textbf{A}(t)$ is the vector potential such that $\textbf{E}(t)=-\partial \textbf{A}/\partial t$ and $\boldsymbol{\pi}=-i\mathbf{\nabla}$) is the interaction term expressed in the velocity gauge. The effective potential $V_{\textrm{eff}}$ of argon used in the calculation accurately fits the argon eigen-energies and -wavefunctions\cite{MullerPotential}. The time-dependent wavefunction $\ket{\psi}$ is expanded on a finite basis of B-splines allowing memory efficient fast numerical calculations with a very large basis set \cite{Bachau2001}:
\begin{equation}
  \psi(\textbf{r},t) = \sum_{l=0}^{l_\textrm{max}}\sum_{i=1}^{n_\textrm{max}} c_{i}^{l}(t)
  \frac{B_{i}^k(r)}{r}Y_{l}^{0}(\theta,\phi),
\label{eq:TDSE_expansion}
\end{equation}
where $B_i^k$ and $Y_l^m$ are B-spline functions and spherical harmonics, respectively. The basis parameters ($l_\textrm{max}$, $n_\textrm{max}$, $k$ and the spatial box size) and the propagation parameters are chosen to ensure convergence. The atom is initially in the ground state (3p) and the electric field $E$ is linearly polarized along the $z$ axis and is expressed as $\textrm{E}(t)=E_0\cos^2(t/\sigma_\textrm{t})\sin(\omega_0t)$ for $|t|<\pi N/\omega_0$, where $\omega_0$ is the central frequency of the laser, $\sigma_\textrm{t}=2N/\omega_0$, and $N$ the total number of optical cycles within the pulse. The simulations are performed for a laser wavelength of 800\,nm and a pulse duration corresponding to $N$=20 cycles. The third-harmonic electric field is expressed as $\textrm{E}_{\textrm{TH}}(t)=\sqrt{R}E_0\cos^6[t/\sigma_\textrm{t}]\sin[3\omega_0t+\varphi]$, where $\varphi$ is the relative phase between fundamental and third-harmonic electric fields and $R$ is the relative intensity of the third harmonic with respect to the fundamental. Ionization yield induced by either a single IR beam or the combination of UV and IR beams were evaluated as a function of both IR and UV peak intensities for $\varphi=0$ and $\varphi=\pi$. Note that with this definition of the electric field, based on sinus functions, constructive (resp. destructive) interference is achieved when $\varphi=\pi$ (resp. $\varphi=0$) unlike the definition used in the figures shown all along the manuscript.
}
\bibliographystyle{nature}

\textbf{Acknowledgements}\linebreak
\footnotesize{This work was supported by the Conseil R\'egional de Bourgogne (PARI program), the CNRS, the French National Research Agency (ANR) through the CoConicS program (contract ANR-13-BS08-0013) and the Labex ACTION program (contract ANR-11-LABX-0001-01). P.B. thanks the CRI-CCUB for CPU loan on its multiprocessor server. The authors gratefully acknowledge V. Tissot and J.-M. Muller for the cell conception and Prof. D. Charalambidis for discussions.}\linebreak
\textbf{Authors~contributions\hfill}\\
\footnotesize{P.B. suggested the idea of the filamentation control by harmonic seeding. P.B. and J.D. performed the simulations. P.B., G.K. and F.B. performed the experiments. All authors contributed equally to the writing of this work.}
\vfill
\end{document}